Preprint. May 2026.

# Rethinking AI Psychosis: Misnomers, Conceptual Limits, and Existential Drift

Kasper Møller Nielsen[1] & Lucy Osler[2]

[1] Center for Subjectivity Research, University of Copenhagen, Copenhagen, Denmark. ORCID: https://orcid.org/0000-0002-3222-1751

[2] Department for Social and Political Sciences, Philosophy, and Anthropology, University of Exeter, Exeter, United Kingdom. ORCID: https://orcid.org/0000-0002-6347-8381

Corresponding author: Kasper Møller Nielsen, Center for Subjectivity Research, Karen Blixens Plads 8, 2300 Copenhagen S, Denmark, tel: +45 26116014, email: kmn@hum.ku.dk

**Keywords**: AI psychosis, phenomenology, existential drift, chatbots, conversational AI, delusion

**Acknowledgements:** We'd like to thank audiences at the Human Abilities Center in Berlin and Egenis in Exeter for their generous responses to some of the core ideas that formed the backbone of this collaborative work. We'd especially like to thank Joel Krueger, Laura Oppi, and Mads Gram Henriksen for helpful comments on an earlier version of this paper.

Funding: This paper was partly funded by a generous research fellowship that Lucy Osler received at the Human Abilities Center in Berlin.

# Rethinking AI Psychosis: Misnomers, Conceptual Limits, and Existential Drift

**Abstract**: There has been a proliferation of media reports about so-called AI psychosis in the last year. Not surprisingly, this has prompted growing academic work on the ways in which AI chatbots such as ChatGPT, Claude, and Replika might aggravate or even induce psychosis, typically understood in terms of users acquiring or maintaining delusional beliefs. Our paper consists of two parts. First, we provide a number of reasons to be sceptical about understanding 'AI psychosis' as a novel psychiatric category. We argue that many of the purportedly new phenomena are better understood through Stompe et al.'s (2003) metaphor of 'old wine in new bottles' and highlight conceptual, nosological, clinical, and social risks associated with the uncritical adoption of this terminology. Second, we develop a positive phenomenological account of what may nevertheless be at stake in sustained human–AI interaction. Rather than focusing primarily on whether AI systems induce, amplify, or sediment delusional *beliefs*, we examine how conversational AI may participate in transforming a person's lived experience of reality itself. We claim that the sycophantic and pseudo-intersubjective nature of AI could lead to what we call "existential drift", whereby individuals may continue to feel rooted in a shared reality through their interactions with AI, while actually becoming entrenched in increasingly private and subjective worlds.

## 1. Introduction

In the past year, reports of so-called *AI psychosis* or *ChatGPT psychosis* have attracted significant media attention (e.g., Hart, 2025; Hill & Freedman, 2025; Tiku & Malhi, 2025). These reports typically describe cases in which prolonged interaction with conversational AI bots (such as ChatGPT, Claude, Gemini, Replika, and character.ai) appear to lead to the exacerbation or development of psychotic symptoms in individuals, usually delusional beliefs. A brief Google search will yield a host of articles on this topic, often with a despairing angle. Scientific research, partially due to its inherently slow institutional nature, is trying to catch up to this media coverage. Presently, there is no systematic empirical research in this domain, only anecdotal evidence, case reports, and commentaries. Work by Morrin et al. (2025, 2026) collects various media cases, and discusses the potentials and dangers of AI in relation to psychosis.

Given the apparent dangers that anecdotal evidence seems to point towards—especially high-profile cases reporting that AI chatbots have encouraged users to commit suicide or homicide—it is not surprising that there is both a sense of growing academic interest and urgency in better understanding how human-AI interactions impact people's mental health. In this context, two key questions have arisen both in the public debate but also in scientific literature. First, whether we should posit AI psychosis as a novel kind of psychosis or mental disorder. Second, whether and how interacting with conversational AI can induce, amplify, and/or sediment psychosis and/or delusional beliefs (e.g., Morrin et al., 2026; Osler, 2026; Østergaard, 2023).

In relation to the first question, we offer a sceptical response. We present some critical remarks about adopting the terminology of either 'AI psychosis' or 'AI-induced psychosis', outlining various conceptual, nosological, clinical, and social risks associated with these terms. We are especially sceptical of the sense of novelty that is often expressed in relation to the term AI psychosis, where people, for instance, denote it as "a new frontier in mental health" (Preda, 2025).[1]

Notwithstanding this negative assessment, we also provide a more positive proposal. Regarding the second question, we agree that more research needs be done both theoretically and empirically to examine the impact of human-AI interactions. As alluded to above, much of the emerging literature on so-called AI psychosis focuses on whether conversational AI systems can induce, amplify, and/or sediment delusional *beliefs*.[2] While these concerns are well-founded, we suggest that they capture only part of the phenomenon. Through a phenomenological lens, we suggest that the more fundamental issue is how sustained interaction with conversational AI may transform a person's lived experience of reality—specifically, their sense of what is shared, objective, and intersubjectively anchored. We consider how conversational AI may play a pseudo-intersubjective role in co-constituting a user's sense of reality. In doing so, we contemplate how interacting with sycophantic conversational AI might lead someone to gradually become unmoored from the shared-social world, whilst increasingly becoming entrenched in private and subjective worlds; a phenomenon we dub 'existential drift'. On our account, existential drift does not necessarily constitute psychosis, nor does it necessarily involve delusional beliefs. Rather, it captures a transformation in the constitution of the experiential background against which beliefs, including delusional ones, emerge.

[1] Hudon and Stip (2025, p. 12) similarly denote human-AI interactions in relation to mental health as "a new frontier in psychiatry".
[2] This reflects a broader tendency to frame the risks of conversational AI primarily in epistemic terms rather than, for example, social, affective, and phenomenological ones (Krueger & Osler, 2025).

The paper proceeds as follows. In section 2, we introduce some key concepts from phenomenological psychopathology to frame our analysis. Section 3 sets out our concerns about treating ‘AI psychosis’ or ‘AI-induced psychosis’ as a novel diagnostic category. In section 4, we consider some pragmatic and social benefits and risks relating to these terms. Finally, in section 5, we turn to what we believe is the novel aspect revealed by the discussions of so-called AI psychosis, namely the particular kind of relationship that a human and conversational AI may develop, and how this has consequences for mental health. Here we build on Morrin et al.’s (2025, 2026) idea that talking to conversational AI may lead to ‘epistemic drift’ and suggest these interactions may lead to ‘existential drift’, a gradual reorientation of a user’s sense of what is real.

## 2. A psychopathological preamble

Our discussion of AI psychosis is rooted in the phenomenological tradition and, as a preamble to the coming analysis, we want to start by introducing some of the foundational concepts in the phenomenological examination of psychosis and delusion.

Phenomenology is a philosophical tradition that aims to uncover the structures of human experience and foregrounds lived experience as the site of study. In doing so, phenomenology seeks to capture the structures of, amongst other things, perception, self, intersubjectivity, and affectivity. A strong vein of interest in this philosophical project has been phenomenological psychopathology—the study of mental illness through careful attention to how psychiatric conditions transform structures of experience. Phenomenological psychopathology has undergone a significant revival in the last 30 years, analysing a wide range of mental disorders including schizophrenia, depression, eating disorders, bipolar disorder, and anxiety and examining a broad spectrum of transformations across self, intersubjectivity, temporality, spatiality, affectivity, perception, and embodiment (see, e.g., Stanghellini et al., 2019). While a key aim of these studies is to better understand how mental illness transforms an individual’s very experience of themselves, others, and the world, many accounts emphasise that examining transformations in mental illness can also work to reveal structures of experience that ordinarily remain invisible—what Gallagher (1997) dubs a *mutual enlightenment*. The study of thought insertion, for instance, makes us consider and reflect on the nature of the self and the different ways the self can be disturbed (e.g., Henriksen et al., 2019). Likewise, work on embodiment may contribute to a more comprehensive understanding of melancholic depression (e.g., Fuchs, 2005).

Much of the current work on AI psychosis that we discuss below invokes the psychiatric concepts of delusion and psychosis. Delusions are usually defined à la the DSM-5 as “fixed beliefs that are not amenable to change in light of conflicting evidence” (American Psychiatric Association, 2013, p. 87). Such definitions often refer to so-called external characteristics of delusions: certainty, incorrigibility, and impossibility (Jaspers, 1997, pp. 95-96). Note, however, that while definitions, such as the one in the DSM-5, might function as a rough characterization of delusion, they come with significant flaws. First, the external characteristics can actually all be disjunctively absent, without its ceasing to be a delusion (Spitzer, 1990). Second, these characteristics make it difficult to demarcate between delusions, overvalued ideas, religious experiences, conspiratorial thinking, superstition, etc. (see, e.g., Haenel, 1983; McKenna, 1984).

Furthermore, despite psychosis being a foundational concept in psychiatry that bears crucial implications for classificatory, treatment, and legal issues, it is poorly understood and defined (Bürgy, 2008). For instance, the concept of psychosis is usually left undefined in the major diagnostic manuals, such as the ICD-11 and DSM-5. The ICD-11 (World Health Organization, 2024, pp. 306, 312) simply lists “features of psychosis”, such as hallucinations or formal thought disorder. The DSM-III (American Psychiatric Association, 1980, p. 367), which only defines the adjective form, ‘psychotic’, reads that it is a “term indicating gross impairment in reality testing”. This is similarly reflected in the ICD-11 (World Health Organization, 2024, p. 161), where it is claimed that schizophrenia, the psychotic disorder par excellence, involves “significant impairments in reality testing”.

The psychoanalytic concept of reality testing (Hurvich, 1970) is, however, problematic and may misconstrue things, especially in relation to schizophrenic psychosis (Parnas, 2015; Sass, 1994). Reality testing usually refers to a capacity of the subject whereby they distinguish between internal mental states (e.g., wishes, fantasies, imaginings) and external perceptual reality. On this view, psychosis represents a failure of this capacity leading the subject to mistake their inner productions for features of the outer world. Moreover, the emphasis on psychosis as impaired reality testing also has a tendency to construe psychosis as something that is either absent or present, like a light being off or on. This, though, overlooks the contextual difficulties in determining psychosis, where it always has to be grasped and understood contextually in relation to the broader sociocultural background (Parnas et al., 2010). This ties back into the difficulties with differentiating delusional beliefs from atypical beliefs, superstition, religious beliefs, etc. While we are not disputing that psychosis involves *a disturbed grasp of reality*, i.e., a displacement in relation to the shared-social world, the concept of reality testing as a definition of psychosis may miss the complexities of psychosis (Henriksen, 2018).

Phenomenological accounts, in contrast, emphasise that a more global alteration in the "awareness of reality" is at stake in psychosis (Jaspers, 1997, pp. 93-94), i.e., the person with psychosis finds herself differently in the world. This transformation is often characterized as a profound disconnection from the shared-social world, leading to a breakdown in mutual understanding with others described as a disturbance of communality (Blankenburg, 1991; Wyrsch, 1951). This sense of communality and belonging is not a set of explicit beliefs or proposition-like statements but instead concerns a pre-reflective attunement to the shared-social world that shapes perception, understanding, and the sense of what is real and what is not (Ratcliffe, 2008). Connected to this estrangement is also the entrenchment of perspective, where the person no longer retains insight into the absurdity of their idiosyncratic viewpoint of the world. This disturbance of communality is often manifestly displayed in beliefs, emotions or behaviour that are discordant with the shared-social world, often displaying a profound alienation from mutual understanding. Specifically in relation to schizophrenia, the psychosis also seems to entail the opening of a solipsistic domain of reality (Henriksen & Parnas, 2014; Nielsen, 2024). Finally, for psychosis to be of psychiatric relevance, it must also concern distress, dysfunction, or suffering (Scharfetter, 2010, p. 212). Thus, what the phenomenological perspective highlights is that construing delusions simply as 'false beliefs' is superficial, and overlooks the transformations to the structures of self, others, and world operative in these experiences.

Needless to say, understanding psychosis and delusions, and their associated experiential structures, are extremely complex issues, and our goal here is not to resolve them. Rather, we simply want to preface these issues as caveats to our analysis of various discussions of psychosis and 'AI psychosis' that may sometimes sideline the complex foundational issues involved in defining delusions and psychosis and doing justice to the lived experience of such phenomena. It is also important to underscore that these problems are not something that simply are salient for the nascent field of AI psychosis, but they are something that haunt psychiatry *simpliciter*. With these reflections in mind, we now turn to the contemporary discussions on AI and psychosis, starting with the terminological issues.

## 3. AI Psychosis: What's in a name?

At the time of writing, some of the most high profiles instances of what is commonly being dubbed AI psychosis include: Jaswant Singh Chail, who in 2021 attempted to assassinate

Queen Elizabeth II with a crossbow after his Replika chatbot “girlfriend” told him his plan was “very wise” and that she was “impressed” he was an assassin (R v Chail 2023); fourteen-year-old Sewell Setzer III, who committed suicide in February 2024 after intensive interactions with a Character.AI chatbot, with his final exchange involving the bot telling him to “please come home” (Garcia v Character Technologies Inc., 2024); Stein-Erik Soelberg, a 56-year-old former tech executive who killed his mother and then himself in Greenwich, Connecticut in August 2025 after ChatGPT repeatedly affirmed his paranoid delusions that his mother was poisoning him (Jargon & Kessler, 2025); and sixteen-year-old Adam Raine, whose parents filed a wrongful death lawsuit against OpenAI in August 2025, alleging ChatGPT acted as a “suicide coach” through thousands of pages of conversations (Betts, 2025). These cases, along with multiple lawsuits filed against Character.AI throughout 2024 and 2025, have driven much of the public concern and media attention surrounding AI, psychosis, and delusions.

Understanding what is going on in cases such as these is a multifaceted problem that especially concerns psychiatry and law. Still, the terms employed in this nascent domain also have profound implications for, amongst others, research, clinical care, software development, and policy. The words we employ partially shape our relationship with reality (Hacking, 1999), and in this regard we should also be mindful of our terms and their associated concepts. Perhaps given the novelty of the field, a standard terminology has not been settled yet. However, especially in the media, the phenomenon is often denoted as AI psychosis or ChatGPT psychosis (see, e.g., Hart, 2025; Kleinman, 2025; Wei, 2025). In the academic literature, similar things are happening. For instance, Yeung et al. (2025) insist on calling it AI psychosis. Analogously, Preda (2025) labels it AI-induced psychosis and dubs it a novel “complex clinical syndrome”.

In our view, the terms ‘AI psychosis’ and ‘AI-induced psychosis’ suffer a number of problems, which we go through below (see also Flathers et al., 2026; Morrin et al., 2026). In broad strokes, the term AI psychosis seems to stipulate a diagnostic entity that is overly inclusive and collapses or lumps psychopathological categories that usually are kept distinct into a single term.[3] Talk of AI psychosis also risks pathologizing behaviour and beliefs that, in some instances, are quite mundane. This may lead us to suggest ‘AI-induced psychosis’ as a more precise term, that, however, by bringing in the concept of induction also opens its own can of worms.

[3] This is especially peculiar in light of a psychiatric climate that has been furiously pursuing splitting of psychiatric categories for the last decades (Frances & Widiger, 2012; Lilienfeld & Waldman, 2004).

### *3.1 AI Psychosis*

#### *3.1.1 The problem of diagnostic over-inclusivity*

It is often unclear what unites the proposed category of AI psychosis. The term is applied to a wide variety of cases, including the aggravation or amplification of a user's pre-existing mental health condition, as well as to instances where it is speculated that the chatbot interaction itself triggered or induced the symptoms. In addition to this, the term also spans a wide range of psychotic experiences, from, for instance, feeling that the AI puts the person in contact with angelic beings to falling in love with their chatbot. This leaves us asking what this term is supposed to denote. Is it a particular form of psychosis? Does it pick out the same content of psychotic or delusional experiences? Is it a suspected common aetiopathology? Or something entirely different? Clarifying these issues is essential for determining whether this in fact is a novel diagnostic category, as well as more generally for establishing consistency within this nascent field of research.

A key concern we have about the term AI psychosis is that it is overly inclusive. The paper by Yeung et al. (2025) is an interesting case in point. They define AI psychosis as: "*The onset or exacerbation of adverse psychological/psychiatric symptoms following intense AND/OR prolonged interaction with (generative) AI systems*" (Yeung et al., 2025, p. 3, italics in original). They introduce this definition, even though they also claim that "LLM-induced psychological destabilisation" is "a more precise term" (Yeung et al., 2025, p. 3). This broad definition, paired with the caveat, underscores the expansive nature of the terms being employed.

This overinclusive definition by Yeung and colleagues seems to suggest that some cases of AI psychosis simply involve the exacerbation of *psychiatric* symptoms. Notwithstanding that defining psychosis is complicated, the alienation from the shared-social world as well as entrenchment of an idiosyncratic perspective, does seem to suggest that psychosis is a *qualitatively* different state. For instance, the exacerbation of worry and insomnia in generalized anxiety disorder (whether or not grounded in interaction with a chatbot) does not amount to psychosis. A patient suffering from these symptoms does not have a disturbed grasp of reality and is often aware of the pathological nature of these experiences, i.e., they retain insight into their condition.[4]

A more fruitful way to grasp AI psychosis and avoid over-inclusivity is perhaps to make finer distinctions. One way to do this is through Jaspers' (1997, p. 58) old distinction between form and content of experiences, i.e., the mode or structure of experience and the object of experience. Following Jaspers, the mode, form, or structure of the experience labels the kind of experience it is, e.g., a perception, an imagination, a memory, or a hallucination. While the content labels what the experience is about. Different forms or modes of experience can all share the same content, e.g., one can imagine, remember, perceive, or hallucinate a white mouse. While, the same form or mode of experience can involve different content, e.g., one can hallucinate a white mouse, a pink elephant, and shadowy figures.

[4] This is not to deny that psychosis may sometimes exhibit this incipient, quantitative increase of symptoms and delusional conviction (cf. Nordgaard & Henriksen, 2026).

One strand of the public and scientific literature of the contemporary debate seems to suggest that the diagnostic category is formed based on the content, i.e., AI psychosis shares the object of involving AI. For example, where individuals believe that their chatbots are conscious, that they are romantic partners, or that they are mediating one's connection to spiritual beings. However, as Jaspers (Jaspers, 1997, p. 59) warns, content may be somewhat irrelevant for understanding what unifies and separates these experiences and associated disorders. For instance, being convinced that the neighbours are spying on you may occur in many different psychotic disorders, such as in delusional disorder, schizophrenia, and Alzheimer's. However, these disorders are different and may require different interventions. Delusional beliefs about AI may, therefore, arise in different kinds of psychotic disorder, and lumping these diagnostic categories based solely on content may be unfortunate, as it may have unhelpful repercussions for psychiatric intervention.

Relatedly, focusing on the form of these disorders might not be very illuminating either. Morrin et al. (2026), in their survey of cases from news media, found that the delusions seem to cluster into: 1) spiritual or messianic delusions, 2) belief in sentient or god-like AI, and 3) romantic or attachment delusions. As further research is conducted, more categories will probably be added. Yet, while the content or delusional object, namely the AI, is novel, the form of the delusion itself is not. We find messianic, grandiose, persecutory, or erotomanic delusions in many kinds of psychotic disorder, both organic and non-organic conditions (i.e., conditions with and without known neuropsychiatric cause), which does not capture the specificity of these different conditions. Instead, what makes these delusions particular—for instance, a persecutory delusion schizophrenic rather than depressive—has more to do with the evaluation of the entire symptom complex, which reveals *a specific tint* of the individual symptoms (Jaspers, 1997, p. 581; Nielsen, 2023).

Whether defined via form or content it seems that the category of AI psychosis risks obscuring rather than illuminating; masking the differences between what may be distinct clinical conditions with different treatment requirements (and perhaps different aetiologies). By focusing on the 'AI' as a unifying marker we risk losing sight of the various underlying psychopathological structures at play.

*3.1.2 An illusion of novelty*

It is also important to remember that the content of delusions is shaped by the surrounding sociocultural environment (“One cannot have a delusion about Jesuits if he has never learned about Jesuits” (Meehl, 1990, p. 35)), while the form of the delusion seems to be stable. Such stability of form led Stompe et al. (2003), more than 20 years ago during a similar technological hype, to caution that what may seem to be a novel delusional form may simply be *old wine in new bottles*. Back then it was, for instance, whether the experience of being a webcam is a novel form of psychosis (Schmid-Siegel et al., 2004). Now the content concerns artificial agents and LLMs. In this regard, we concur with Carlbring and Andersson (2025, p. 1) that “[w]hile alarming, [AI psychosis] is most likely not a new unique diagnostic entity”.

From a historical vantage point, we should also note that delusions often involve technology and machinery. Morrin et al. (2026, p. 3) also point to this stability of content, emphasising the presence of technology in delusional content in the last 100 years (drawing on Higgins et al., 2023). Indeed, we can actually trace this back further; one of the oldest records we have of schizophrenia, from 1810, namely the account of James Tilly Matthews, involves delusional ideas about being influenced by a mechanical “air loom” (Haslam, 1810) and, in 1885, Kandinsky recounted his patient Perewalow’s experiences of being tormented by what he called *Stromisten* (currentors), that inserted disturbing images into his head with a special machine (Kandinsky, 1885, pp. 78-79).

We, of course, recognise that the construction of new diagnostic categories when the underlying causes are unknown remains one of the thorniest issues in psychiatry. Indeed, these challenges are not unique to the study of AI and mental health; they haunt psychiatry *simpliciter*. Nevertheless, rather than assuming novelty, we believe progress is best made by clarifying these issues conceptually while simultaneously conducting systematic empirical work. Only by situating AI within this broader psychiatric context can we address the genuine impact these technologies have and assess their novelty.

### *3.2 AI-induced psychosis*

AI-induced psychosis is another term being employed. This may on first blush seem more appropriate, as it shifts the emphasis to *the generative aspect* of psychosis. In other words, it is intended to point to the role that AI may play in the development of psychosis or the induction of psychosis. This, then, might be viewed as a mechanistic label. What is being denoted as 'novel' is not a new psychiatric category *per se*, but a specific, interactive pathway through which psychotic conditions are kindled or exacerbated. However, as we will argue, while this shifts the focus toward the interesting dynamic role of the AI, the terminology still faces some difficult hurdles.

It is also first worth noting that some uses of 'AI-induced psychosis' suffer from the same drawbacks as identified above. For instance, Preda (2025), adopts a very broad definition of AI-induced psychosis as: "a complex clinical syndrome in which psychotic symptoms overlap with mood changes, limited insight, poor judgment, neurovegetative symptoms, and behavioral changes". Similar to Yeung et al. (2025) above, we see a hodgepodge of symptoms and behaviours that are quite unspecific and do not allow for a differential diagnosis.

*3.2.1 A question of causality*

Notwithstanding that some uses of AI-induced psychosis suffer from the diagnostic over-inclusivity we addressed previously, the issue we want to draw attention to is the causal weight it places on the technology itself. While there has been a lot of news-coverage of high-profile reports of 'AI psychosis', given the significant adoption and use of conversational AI, there might already be some reason to suppose that the idea of AI as a primary "inducer" or "catalyst" of psychotic symptoms is misleading (cf. Flathers et al., 2026). If AI interaction were capable of inducing psychosis *de novo*, we might expect to see significantly higher rates of clinical incidents. Instead, it might be supposed that the human-AI interaction seems to have the potential to kindle or aggravate pre-existing mental health issues—and relatedly, that perhaps these individuals also had vulnerabilities that made them seek out more intense interactions with a chatbot in the first place (Morrin et al., 2026, p. 6).

This brings us to the problematic assumptions regarding "induction" and "causation" inherent in the term AI-induced psychosis. Contemporary diagnostic manuals purport to be atheoretical and descriptive, since the aetiopathology of mental disorders is currently unknown. Even in cases where a cause is posited, such as in post-traumatic stress disorder and substance-induced psychosis, the relationship between causal agent and clinical syndrome is more complicated than often presented, and it does seem to involve certain predispositions in addition to a causal agent (Bramness et al., 2024; Paris, 2022). The issue with the concept of AI-induced psychosis is that it posits a seemingly straightforward link between interaction with an AI and psychosis. At its most egregious, it might be equivalent to talking about 'interpersonal conflict psychosis' or 'job loss psychosis'—which is, at best, a caricature. Causation in psychiatry is never this straightforward (Kendler et al., 2011; Woodward, 2008).

Moreover, at this point in time, whether “LLMs are capable of inducing a persistent state of psychosis in somebody with no history and without excessive risk factors” is an open question (Morrin et al., 2025, p. 14); and, given the difficulties just identified, may well remain open.

*3.2.2 Technological folie à deux*

Another way researchers have attempted to frame the idea that AI induces psychosis is through the lens of *technological folie à deux* (Dohnány et al., 2026). Classically, *folie à deux* describes a psychiatric phenomenon where a “genuine” psychotic individual (the inducer) transmits their delusions to a closely associated non-psychotic person (the inducee) (Lasègue & Falret, 1964; World Health Organization, 1992).[5] In the case of AI and psychosis, the emphasis is (supposedly) shifted, where the AI is the inducer and the person is the inducee. Besides this being a peculiar anthropomorphization of attributing psychosis to the AI, it also misses Lasègue and Falret’s (1964, p. 9) classical distinction between “insanity” in the inducer and “absurdity” in the inducee. The AI is not insane, and the person is not only absurd. Thus, the picture of folie à deux in the case of AI may be more figurative than accurate.

In the end, given the absence of systematic research, we do not know what most of the people in the cases diagnostically are suffering from. The cases collected by Morrin et al. (2026) suggest a mixed picture of various kinds of non-organic psychosis, such as schizophrenia, bipolar disorder, and acute and transient psychotic disorder. Of interest for further research might be the fact that many of these cases seem to suffer from acute and transient psychotic disorder or what the French tradition called *bouffée délirante*, i.e., the persons seem to experience predominantly delusions with an acute onset, which runs the course of a few weeks, perhaps even days, with no apparent precipitating event or stressful circumstances, and makes a rapid and good remission (Pichot, 1986; World Health Organization, 2024, p. 180). This is of course a hypothesis and something that will have to be borne out in future research. Nevertheless, as return to below, what the focus on induction perhaps most clearly underscores is that what we should be paying attention to is the dynamics of human-AI interactions.

[5] In the latest diagnostic manuals, DSM-5 and ICD-11, *folie à deux*, or induced delusional disorder (ICD-10) or shared psychotic disorder (DSM-IV) as it was also called, has disappeared as a standalone diagnostic category, and thus is not uncontroversial.

## 4. A pragmatic interjection on the term 'AI psychosis'

It is worth finishing our discussion of labels on a pragmatic note. Here we consider the potential risks that such terminology could lead to over-pathologisation and stigmatization but also recognize the potential power of provocative terms like AI psychosis.

### *4.1. Over-pathologisation and stigmatisation*

Beyond the conceptual confusion that the term AI psychosis elicits, it also raises concerns of the over-pathologisation and stigmatisation of user behaviours and experiences. We are seeing a proliferation of experiences—from intense romantic attachments to the adoption of false beliefs via "AI hallucinations"—that may appear new and (to some) even worrisome without necessarily indicating a psychiatric disorder.[6] These cases certainly point to very real, and sometimes very significant, affective, social, and epistemic impacts that come with human-AI interaction (a point we return to below). However, they also point to a number of potential risks that come with the proliferation of the term AI psychosis.

First, a blurry-edged label like AI psychosis might risk being applied to cases that we do not want to pathologise—an additional concern of over-inclusivity. As noted in section 2, psychosis is not merely the presence of "false" or "odd" beliefs. Instead, it involves an alienation or estrangement from the shared-social world, as well as a rigidity of convictions and perspectives. Users may hold idiosyncratic beliefs that have been enforced or even introduced by conversational AI or develop intense romantic attachments for their chatbots yet remain anchored in collective reality.

Second, we might risk overlooking that altering one's beliefs or even falling in love with an AI chatbot are not only predictable outcomes of engaging with AI chatbots but maybe even be appropriate responses to systems explicitly engineered to produce the social engagement, affective attachment, and trust of their users. Just as we do not pathologise a viewer for mourning a fictional movie character—recognising that the narrative was specifically engineered to evoke such feeling—perhaps we should be cautious labelling emotional attachments to AI as 'psychosis'.[7]

[6] For example, we are increasingly seeing a wide range of users reporting intense emotional, and even romantic, attachments to their chatbots; sometimes grieving deeply when system updates lead to their AI companions being "lobotomised" or "killed". We have also seen significant concern being raised about the overly agreeable (sycophantic) tone of chatbots leading to people's own beliefs and narratives being uncritically affirmed (Osler, 2026, forthcoming). As well as significant discussion about the epistemic risks of so-called 'AI hallucinations' or 'AI bullshit' that might lead to users adopting false beliefs about the world (Fisher, 2024; Hicks et al., 2024; Krueger & Osler, 2025), which to some eyes may indeed appear to be delusional.

Finally, there is the concern of stigmatisation. On the one hand, users with these attachments may be loath to talk about their experiences, or seek help if needed, out of fear of being labelled psychotic or delusional. Stigmatisation could have the unfortunate effect of preventing early invention, where required, if users fear discussing their experiences with family members, friends, or clinicians—precisely the people who might be best placed to notice concerning patterns or vulnerabilities. On the other hand, we also should be cautious about creating stigma and shame around emotional attachments for chatbots more generally. As Danaher (2019) stresses, we should adopt some epistemic humility when it comes to thinking about how and why users might develop relationships with chatbots and not fall into a knee-jerk pathologisation of such experiences.

*4.2. The political power of terminology*

While we have emphasised that there is, as yet, no robust empirical evidence to suggest that AI psychosis in fact is something entirely novel in the psychiatric landscape, it can hardly escape our notice that powerful and seductive AI chatbots are being rolled out into society subject to little regulation. Indeed, how users respond to these AI systems is part and parcel of the research development process of tech companies, many of whom explicitly endorse the mantra ‘Move fast and break things’. A mantra we should be concerned about when those ‘things’ could include our own mental well-being.

Thus, one reason to keep using terms like AI psychosis is to raise public and political awareness of the potential harms that might arise through human-AI interactions. The eye-catching term could serve a crucial critical function in the public sphere, forcing both conversation about and research into the impact of increasingly companion-like AI systems. Moreover, it works to centre AI companies like OpenAI, Meta, xAI, Google, Anthropic, and Replika and their responsibility for developing and deploying new technologies without requisite safeguards. In this sense, it is important to recognise the powerful rhetorical force that a term like AI(-induced) psychosis has for ongoing questions about regulation, policy, and accountability; working to put pressure on companies and developers to take responsibility for and implement safety measures around their products.

[7] For accounts that frame our experiences of AI chatbots in fictionalist terms, see, e.g., Krueger and Osler (2022); Mallory (2023); Krueger and Roberts (2024); Friend and Goffin (2025).

The question of accountability is pressing. AI companion systems are not neutral tools but designed artefacts, shaped by choices about tone, persona, memory, and responsiveness. When companies optimize for engagement, knowing that emotional attachment and sycophantic affirmation drive user retention, they are making design decisions with potential mental health consequences. A provocative term like AI(-induced) psychosis, whatever its conceptual limitations, rightly draws attention to the way that AI can be an active participant in shaping users' experiences. In the absence of robust regulatory frameworks, evocative terminology may be one of the few tools available for holding powerful actors to account.

Indeed, it did look like public uproar around the excessively sycophantic nature of chatbots like ChatGPT, criticised for providing no friction to users' beliefs, and even suicidal ideations, did lead to OpenAI explicitly toning down the agreeableness of ChatGPT5. But many users complained that they found this model cold and informal. This resulted in OpenAI dialling the friendly tone back up again weeks later. This rather suggests that if sycophancy drives engagement, and at the end of the day engagement drives revenue, then even a term like AI-induced psychosis can be shrugged off.

## 5. Human-AI interactions: pseudo-intersubjectivity, affirmation, and existential drift

Our critical evaluation of the terms AI psychosis and AI-induced psychosis should not be taken to suggest that contemporary human-AI interactions are psychiatrically uninteresting or even benign. On the contrary, while we remain sceptical about the idea of postulating a novel diagnostic category, the anecdotal evidence collated by those such as Morrin et al. (2026), should prompt closer investigation of the structure of human-AI interactions and how they shape our experiences and mental wellbeing. There are many avenues for exploration in this domain, asking and analysing the role that AI companions might play, both positively and negatively, in the shaping of our wellbeing, and considering a wide breadth of conditions that AI interactions might shape from anxiety to depression to eating disorders to psychosis.

Our specific interest here is in the way in which sustained interactions with conversational AI systems can come to structure a user's experience of reality itself. To capture this dimension, we suggest that a phenomenological analysis of human–AI interaction is instructive.

### *5.1 Sycophancy and epistemic drift*

A key mechanism that has emerged in recent discussions of AI psychosis is sycophancy—the tendency of chatbots to agree with users, calculating and confirming their perspectives while refraining from challenging them (Morrin et al., 2026; Osler, 2026). Importantly, this sycophancy is not accidental but arises from the way these systems are designed and trained. Chatbots typically rely on reinforcement learning from human feedback, optimizing for responses that users rate positively and that sustain engagement over time. The result is that conversational chatbots typically align themselves with the user's expressed views, interpretations, narratives, and evaluations; leading to a generation of AI chatbots that are particularly agreeable, non-confrontational, and often outrightly flattering.

Where friends, family members, or clinicians might introduce alternative perspectives, friction, or even doubt, in the face of certain views, interpretations, or experiences of another, chatbots risk reinforcing, and potentially amplifying, overvalued ideas or delusional beliefs by adapting to the user's tone, narrative, and interpretive framework. As Morrin et al. (2026, p. 4) observe, this can result in a "progressive weakening of confidence in shared reality or accepted knowledge structures", which they dub "epistemic drift" (Morrin et al., 2026, supplementary appendix, p. 11). Where a user increasingly places their trust in the AI bot, and the outputs it generates diverge from accepted knowledge (either due to its own inaccurate outputs or through alignment with the user's own delusional construal of reality), we can see how a user's belief system might become cut adrift from the shared world with others. Note that this might occur in a number of ways, including trusting what the bot tells you more than what external sources say, feeling that the belief system constructed with the bot is more reliable than consensus views, or experiencing mainstream knowledge as irrelevant compared to the private understanding you have developed supported by your AI companion.

Several studies have attempted to model or simulate these dynamics. Yeung et al. (2025), for instance, demonstrate how repeated affirmation by a chatbot can give rise to what they describe as "delusion reinforcement loops". Similarly, Dohnány et al. (2026) highlight feedback loops between user beliefs and AI responses that may amplify initial epistemic distortions, resulting in what they describe as echo chamber dynamics. Across these accounts, the central concern is the absence of epistemic friction; the corrective resistance ordinarily provided by other subjects within a shared-social world.

Certainly, the attention given to sycophancy and belief reinforcement is warranted, and we do not wish to downplay the epistemic risks involved. However, these accounts predominantly frame the dangers of sycophancy in cognitive or doxastic terms, focusing on how chatbots may confirm, entrench, or amplify particular *beliefs*. What tends to fall out of view on this picture is the more fundamental phenomenological shift that may occur through sustained affirmation—namely, how such interactions may reshape a user's sense of reality itself. Affirmation does not merely operate at the level of individual propositions; it may also function as a form of existential validation, stabilizing a particular way of experiencing the world and one's place within it. It is this deeper transformation, we suggest, that calls for a phenomenological analysis, and the occurrence of what we call 'existential drift'.

### *5.2 Human-AI lifeworlds and existential drift*

Unlike epistemic drift, which focuses on the adoption or enforcement of false or atypical beliefs, existential drift, as we define it here, involves a gradual reorientation of a user's sense of what is real and ordinary. In other words, existential drift concerns a transformation on an existential, not only doxastic, level in relation to others and the world. It creates a rift between the person and the shared-social world, whilst simultaneously disclosing reality in a new way, thus stabilizing a particular, often idiosyncratic, perspective on the world. To bring out what we see as characteristic of existential drift, we first need to introduce the phenomenological conception of the lifeworld.

A key insight from phenomenology is the idea that the world is not simply an objective, physical world 'out there'. Rather, the world is something we find ourselves in, as something already meaningful, familiar, and taken for granted (Husserl, 1989, p. 195). The way the world opens or discloses itself to us is dependent on our beliefs and concerns, as a dynamic interplay between self and world. Chairs and tables, for instance, offer action possibilities and meanings thanks to our embodied and social-cultural being. Our "lifeworld", as Husserl (1970) dubs it, is not encountered as a neutral backdrop but as an environment already structured by expectations, goals, and forms of engagement.

Note, though, that the idea that the world is constituted through our own embodied and practical concerns is not intended to leave us with a solipsistic picture, where we all inhabit our own individual lifeworlds. The lifeworld is fundamentally pervaded by others and intersubjectivity. This is also why Husserl (1970, p. 209) describes the lifeworld as "the world for us all": "the world exists not only for isolated men but for the community of men; and this is due to the fact that even what is straightforwardly perceptual is communalized" (Husserl, 1970, p. 163). In the perception of objects, such as chairs and tables, others are co-present, or appresented, and it is in this encounter with others that the world gains its objectivity. The presence of others whose perspectives interlock with and supplement, and sometimes correct, one's own perspective forms what Husserl (1989, p. 213) calls "a relation to a nexus of persons" which grants us the common, objective world.

It is in the context of the lifeworld as intersubjectively constituted that we believe the concept of existential drift comes into view. A genuine other brings their own perspective, their own grip on a world that is also theirs. It is precisely this alterity that allows intersubjective constitution to anchor us in a common reality: the other's perspective provides friction, correcting, supplementing, and sometimes resisting my own. A conversational AI, by contrast, has no perspective of its own. It does not inhabit the world and has no stake in how things are. Its responses are generated to align with the user's input, not to express an independent view. And yet, AI chatbots can simulate dialogical exchange in ways that lead users to experience them as ‘other-like’ (Coeckelbergh, 2011; Krueger & Roberts, 2024; Osler, 2026). Through their interactive, social, and affective character, AI companions can engender something that exceeds the mere confirmation of propositional beliefs; they can produce the sense of being seen and recognised, of having one's experience of the world affirmed by another. The chatbot behaves as if it shares our situation, and we pre-reflectively take this up. Their affirmations carry lifeworld-constituting power even though they lack the genuine alterity that would make that constitution robust.

Importantly, for most users, this lack of genuine alterity may be of little consequence. The AI functions as just one ‘voice’ in a wider network of human relationships.[8] But for individuals whose connection to the shared-social world is already tenuous, such as those perhaps already struggling with mental health or social isolation, this situation may be different. Here, the AI may come to play a disproportionately influential role in the constitution of the user’s world, i.e., by providing relational validation without friction, this conversational partner may have a disproportionate influence on the user’s sense of reality and place in the world.

Over time, then, interacting with an AI companion can lead to a gradual reorientation of one’s sense of what is real. This orientation does not simply concern doxastic attitudes, but one’s place in the world as such. The person may become gradually estranged from what is common and ordinary, and instead discloses something idiosyncratic or unorthodox about the world. Without genuine alterity, the chatbot's persistent affirmation does not simply sustain a user’s (potentially delusional) belief, it progressively naturalises a particular way of experiencing things. This is existential drift.

While speculative, we think this existential drift could occur in, at least, two ways. On the one hand, this might lead to an experience where the user gradually grows more sceptical about ordinary things and instead becomes entrenched in their particular perspective on reality. In other words, the existential drift is itself experienced, with the user feeling a tension between the world they inhabit with their AI companion and the world around them. On the other hand, the pseudo-intersubjective role of the chatbot could make the drift particularly insidious. As the AI continues to affirm and “understand” the user, it could obscure the experiential markers of isolation that usually signal a detachment from shared reality, allowing the individual to feel anchored in a world shared with their AI even as they drift away from the common intersubjective world. The taken-for-granted character of the lifeworld is quietly reconstituted around the user's framework, stabilised by pseudo-intersubjective confirmation. In other words, the user undergoes an existential drift, without experiencing it as such.

[8] Note, too, that it is AI companions’ ability to present as a pseudo-other that might provide certain affective benefits such as leading to a person feeling heard and seen. This might be particularly important for those who lack this experience of recognition in their day to day lives and might play a role in alleviating loneliness (Peng et al., 2025). But it is also precisely this capacity that gives AI companions their power to reshape a user's relation to reality. Their danger, in other words, is parasitic on their appeal.

Now it is worth noting that we do not take existential drift to be something that only occurs through the course of a human-AI interaction, nor is it necessarily pathological. What we have described is, in certain respects, akin to the process of echo chambers, conspiracy groups, or cults, where a particular, unorthodox perspective on the world is affirmed, whereas more typical ways of understanding the world often are understood as a mirage or illusion (cf. Schmidt-Boddy et al., 2026). Unlike echo chambers or conspiracy groups, however, where a genuine community exists, albeit with atypical beliefs, practices, and convictions, the lifeworld constituted by the human-AI interaction is quite different. There is no genuine community in this interaction, but it simply concerns the person and a pseudo-interpersonal partner. Given the sycophantic and affirming nature of chatbots, in a sense, one could claim that they come to constitute a self-world.

### *5.3. Existential drift and psychosis*

We believe that existential drift may be a factor operating in the development of psychosis in the human-AI interactions. Still, we think it likely that certain predispositions and interacting factors in all likelihood have to be present for the person to develop psychosis. Where genuine psychosis starts and severe existential drift ends, is, of course, difficult to demarcate. A sliding out of the shared-social world, as well as the formation of an idiosyncratic perspective on the world, is characteristic of psychosis in general. Still, psychosis is also characterized by an entrenched perspective on the world, where the person can no longer suspend this particular, idiosyncratic perspective (Blankenburg, 1991). They can no longer see the world as others would. In other words, they can no longer recognize the absurdity of this existential drift (i.e., they no longer retain insight).

The concept of existential drift highlights why several authors seem to believe that chatbots may have a profoundly negative impact on reality orientation and may contribute to the development of psychosis. Authors in the literature usually conceptualize this existential drift as a gradual development of impaired reality testing. Carlbring and Andersson (2025, p. 2), for instance, underscore that chatbots seem to “often favor agreement over reality testing”. This is the same idea that we find in the studies that try to simulate these human-AI interactions that supposedly lead to dysfunctional interactions, potentially spiralling into psychosis (Dohnány et al., 2026; Yeung et al., 2025).[9] But, as we discussed above, we do not believe that reality testing always is appropriate to capture psychosis. In our view, psychosis entails an unmooring from what is ordinarily trusted, understood as an attenuated grasp of reality, rather than an impaired ability to ‘check’ what is real and what is not. In this regard, epistemic drift—developing atypical beliefs—alone is inadequate to understand the existential drift also operative in psychosis. Notwithstanding these conceptual squabbles, we do see why discussions of reality testing and AI appear to be crucial for understanding the phenomenon of AI psychosis.

The importance of these discussions are, for instance, reflected in work by Hudon and Stip (2025, pp. 8-9), where they argue that the sycophancy of chatbots can be seen as “an inversion” of cognitive behavioural therapy for psychosis (CBTp), i.e., rather than challenging delusional beliefs (which is characteristic of CBTp), chatbots may reinforce beliefs. This ties into their peculiar claim that interaction with a chatbot may lead to a kind of schizophrenic autism: “AI-psychosis represents […] a technological variant of phenomenological autism: a retreat into a world interpreted, validated, and enclosed by algorithmic dialogue” (Hudon & Stip, 2025, p. 9). Their idea seems to be that chatbots may pull the person away from the shared-social world and into an idiosyncratic perspective of reality, akin to the Bleulerian conception of schizophrenic autism as a withdrawal from reality into an inner, rich fantasy life (Bleuler, 1950, pp. 14, 63). This idea by Hudon and Stip is basically what we understand as existential drift (perhaps developing into psychosis). Nevertheless, calling it “phenomenological autism” (which is a misnomer) is unfortunate since it misconstrues the nature of schizophrenic autism (Blankenburg, 1987; Henriksen et al., 2025). As we emphasized throughout, psychosis is not one thing, and especially schizophrenic psychosis (and its pre-psychotic alienation from the shared-social world) is something quite specific that cannot simply be equated with existential drift. This relates back to the problem with over-inclusivity above. We have to be mindful of the specific phenomenological differences in these states and conditions. Nevertheless, the suggestion that human-AI interactions may have a profound impact on the person's position in the lifeworld is, we think, conceptually helpful.

We believe that the concept of existential drift brings out the profundity of human-AI interactions, and how—despite our pessimism concerning terminology—AI might have a serious impact on our mental health. This pseudo-intersubjective relationship, where the person may become gradually unmoored from the shared-social world, whilst sustaining a (false) sense of communality, is perhaps something novel in the psychiatric landscape and something we have to consider carefully.

[9] Morrin et al. (2025, p. 8), in contrast, somewhat hesitantly, suggest that AI might have the potential to “support with reality-testing”; it might function as “a reality checking tool”

## 6. Conclusion

Here we have attempted to provide a sceptical yet constructive contribution to the emerging debate on AI psychosis. Specifically, we argued for a number of reasons to be sceptical about novel terms such as ‘AI psychosis and ‘AI-induced psychosis’—especially in relation to the assertion that these constitute novel diagnostic categories. We believe that there are good reasons to suspect that this might simply be a case of old wine in new bottles. We do, however, also acknowledge that parts of this discussion will have to be borne out empirically. Nevertheless, we do insist that an uncritical adoption of this new terminology may come with various conceptual, nosological, clinical, and social complications.

In the second, more positive part of our paper, we developed a phenomenological perspective on what might be at stake when thinking about how interacting with AI chatbots might shape and influence our experience of reality. As we argue, the transformations and alterations that might occur for individuals engaging in these human-AI interactions not only concern their beliefs and doxastic attitudes. Rather, conversational AI might contribute to a transformation of the person's sense of reality and place in the world. We call this existential drift. In particular, we see this existential drift as operative in cases where the sycophantic, pseudo-intersubjective nature of AI can lead to a gradual unmooring from the shared-social world, while the individual perhaps still feels anchored in a community. Existential drift and the power that chatbots may have in contributing to this effect might in fact be a novel aspect of the human-AI interactions and their relationship to mental health.

To understand what is actually going on in these relationships between persons and chatbots, we believe that it is worthwhile to return to the phenomenon itself, which motivates further phenomenological research. In particular in relation to mental health and how human-AI interactions might, for better or worse, alter a person’s lived experiences of the world, themselves, and others. This, of course, demands further empirical and theoretical work; especially as we are perhaps only beginning to see the impact these interactions may have on our lives and mental health.